\documentclass[11pt,a4paper]{article}
\pdfoutput=1
\usepackage{jcappub}
\usepackage{color}

\usepackage{amssymb}
\usepackage{amsmath}

\usepackage{bm}

\newcommand{\be}{\begin{equation}}
\newcommand{\ee}{\end{equation}}
\newcommand{\ba}{\begin{eqnarray}}
\newcommand{\ea}{\end{eqnarray}}

\title{Measuring the Homogeneity of the Universe Using Polarization Drift}

\author[a,b]{Raul Jimenez,}
\author[c,d]{Roy Maartens,}
\author[a,e]{Ali Rida Khalifeh,}
\author[f]{Robert R. Caldwell,}
\author[g]{Alan F. Heavens,}
\author[a,b]{Licia Verde}

\affiliation[a]{ICC, University of Barcelona, Marti i Franques, 1, E08028 Barcelona, Spain.}
\affiliation[b]{ICREA, Pg. Lluis Companys 23, Barcelona, 08010, Spain.} 
\affiliation[c]{Department of Physics \& Astronomy, University of the Western Cape, Cape Town 7535, South Africa.} 
\affiliation[d]{Institute of Cosmology \& Gravitation, University of Portsmouth, Portsmouth PO1 3FX, United Kingdom.}
\affiliation[e]{Dept. de  Fisica Cuantica y Astrofisica, University of Barcelona, Marti  i Franques 1, E08028 Barcelona, Spain.}  
\affiliation[f]{Department of Physics \& Astronomy, Dartmouth College, 6127 Wilder Laboratory, Hanover, New Hampshire 03755, USA.}
\affiliation[g]{Imperial Centre for Inference and Cosmology, Imperial College London, Prince Consort Road, London SW7 2AZ, UK.}

\emailAdd{raul.jimenez@icc.ub.edu; roy.maartens@gmail.com; ark93@icc.ub.edu; robert.r.caldwell@dartmouth.edu; a.heavens@imperial.ac.uk; liciaverde@icc.ub.edu}

\abstract{We propose a method to probe the homogeneity of a general universe, without assuming symmetry. We show that isotropy can be tested  at remote locations on the past lightcone by comparing the line-of-sight and transverse expansion rates, using the time dependence of the polarization of Cosmic Microwave Background photons that have been inverse-Compton scattered by the hot gas in massive clusters of galaxies.  This probes a combination of remote transverse and parallel components of the expansion rate of the metric, and we may use radial baryon acoustic oscillations or cosmic clocks to measure the parallel expansion rate. Thus we can test remote isotropy, which is a key requirement of a homogeneous universe. We provide explicit formulas that connect observables and properties of the metric.}


\begin{document}

\maketitle


\section{Introduction}

Isotropy and homogeneity of the background are basic assumptions of the current standard model of the Universe. Within this expanding background, structure formation proceeds via small perturbations with a possible origin in quantum fluctuations of the vacuum. The homogeneous standard cosmological model is a simple, predictive model that successfully accommodates all observations up to now \cite{Planck18}. However, we should probe the foundations of this model as far as possible in order to understand if it holds and if new physics has not been dismissed because of our assumptions (see e.g. the reviews in~\cite{2010CQGra..27l4008C,2011RSPTA.369.5115M,2012arXiv1204.5505C}). 

Isotropy is well confirmed by observations of the cosmic microwave background (CMB): the temperature of the CMB in its rest-frame shows isotropy at better than one part in $10^4$ \cite{Planck18}. Homogeneity, on the other hand, is {\em not} established by observations of the CMB and the large-scale galaxy distribution -- {\em we cannot directly observe homogeneity,} since we observe down the past lightcone, recording properties on 2-spheres of constant redshift and not on spatial surfaces that intersect that lightcone. What these observations can directly probe is isotropy about the observer. In order to link isotropy to homogeneity, we have to assume the Copernican Principle, i.e. that we are not at a special position in the Universe. The Copernican Principle is not observationally based; it is an expression of the intrinsic limitation of observations from one spacetime location\footnote{Nothing precludes that we are at a peculiar location. In fact, we are in the middle of a void with two massive galaxies, Andromeda and the Milky Way; this in itself is very peculiar \cite{Fattahi}.}.

Of course, there is a rich literature of inhomogeneous cosmological models. In particular, void models aim at explaining the current acceleration of the Universe without the need of a cosmological constant (see e.g.~\cite{2012arXiv1204.5505C} for a review) and while these models suffer from difficulties to fit all observations (e.g.~\cite{2012PhRvD..85b4002B,Roland2012}), it is not ruled out that some better models could be built in the future. 
It is therefore important that we develop direct tests of homogeneity that do not assume the background spacetime.
Checking whether galaxy number densities approach homogeneity on large enough scales (for recent work, see e.g.~\cite{Scrimgeour:2012wt,Laurent:2016eqo,Park:2016xfp,Ntelis:2017nrj,Goncalves:2017dzs}) is based on assuming a Friedmann background and is therefore a {\em consistency} test, not a direct test of homogeneity.

Direct tests of homogeneity need to access the {\em interior} of the observer's past lightcone.  In the case of galaxy surveys, Bonnor and Ellis \cite{1986MNRAS.218..605B} formulated a conjecture about  thermal histories in separated regions of the Universe. The conjecture was developed by some of us~\cite{HJM11} into a direct  probe of homogeneity, by using the ``fossil" record (star formation history) of galaxies. This was then applied to find the first direct constraint on inhomogeneity in a galaxy survey, using the fossil record of SDSS galaxies~\cite{fossil}. The fossil record from the star formation history of galaxies was used as a proxy to probe inside the past lightcone, and led to constraints at the $\sim 10\%$ level on any deviation from the homogeneous Friedmann metric. While the fossil record provides already very interesting constraints, it is not a direct probe in the purest sense, as it uses a proxy to probe the metric. Furthermore, it is always useful to have several probes of the same measurement, so as to minimize possible systematic uncertainties. In this work we will present a method that uses photon geodesics to probe the metric, which is a more direct probe of homogeneity.

In the case of the 
CMB,  the thermal Sunyaev-Zeldovich  effect  probes the remote CMB monopole as seen from the observed galaxy cluster, and thus can provide a direct test of remote isotropy and hence of homogeneity, as pointed out by~\cite{1995PhRvD..52.1821G} (subsequently used to test void models by \cite{2008PhRvL.100s1302C,Moss:2010jx,Caldwell:2013fua}). Similarly, the kinetic SZ effect probes the remote dipole and was used by \cite{Zhang:2010fa} to test void models.  The kinetic SZ can be used as a probe of isotropy inside the past lightcone, and thereby as a probe of homogeneity, if we can observe photons that are multiple-scattered or if we can observe the CMB over an interval of cosmic time~\cite{Clifton:2011sn}. In fact, the long time baseline is critical to our plans: more spacetime  geometry can be accessed by a patient cosmologist \cite{Stebbins:2012vw}.

Polarization of the  SZ effect provides further important tests.  The polarized thermal SZ probes the remote quadrupole, allowing in principle for a reduction in cosmic variance in a perturbed Friedmann universe \cite{Kamionkowski:1997na,Seto:2000uc}. (See \cite{Liu:2016fqc,Terrana:2016xvc,Deutsch:2017cja,Deutsch:2017ybc}
 for recent work on reducing cosmic variance in perturbed Friedmann models via the kinetic and polarized thermal SZ effects.)
 
In this paper, we propose a new method to directly probe homogeneity, based on  changes of the
polarization of CMB photons  generated by inverse Compton scattering of CMB photons off hot electrons in massive  (proto)-halos, and the radial expansion history of the Universe. The new method enables a test of isotropy at remote positions on our past lightcone -- a key test of homogeneity.

In Sec. 2 we  review the description of  expansion rates in a general spacetime. This is a necessary step because to test homogeneity we have to work with space-time metrics that do not rely on homogeneity.
For the same reason,  in general cosmological spacetimes (i.e. without assuming a background or any large scale
symmetries) we cannot describe polarisation as in homogeneous spacetime. This is  presented  in  Sec. 3. In Sec. 3 we also  describe the effect of scattering (by hot  electrons) of  CMB photons in generic metrics and the signature that inhomogeneities  leave on the polarisation signal. Finally in Sec 4 we present an estimate of the observations needed to constrain homogeneity with the
method developed above. We conclude in Sec. 5.

\section{Expansion rates in a general spacetime}

\begin{centering}
\begin{figure}[t]
\hspace*{2.5cm}
\includegraphics[width=.7\columnwidth,angle=0]{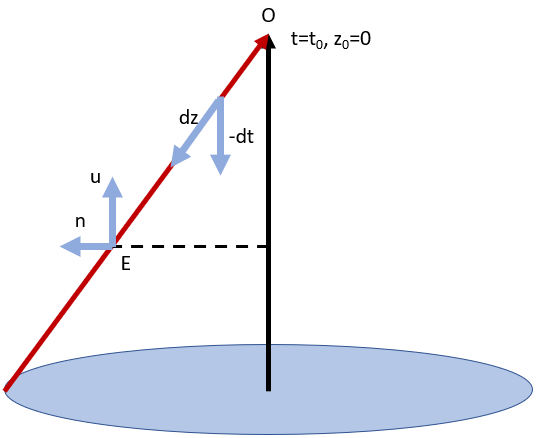}
\caption{Schematic of the lookback time in a general spacetime.}
\label{fig-lb}
\end{figure}
\end{centering}

Let us first recall how to reason in general spacetime metrics. The most efficient way is to use covariant language.
A distant object, with worldline ${\cal E}$, emits photons at event $E$ that we observe with redshift $z_E$ at event $O$ on our galaxy worldline ${\cal O}$. (See Fig. \ref{fig-lb}.)
In order to compare the intrinsic properties of ${\cal E}$ and ${\cal O}$ at the same proper time, we need to compute the look-back time $t_O-t_E$, where
$t$ denotes proper time along galaxy worldlines. This is straightforward in a Friedmann model -- but we cannot assume the geometry of the spacetime if our aim is to test directly for homogeneity. So we need to compute the look-back time in a covariant way, valid in a general spacetime~\cite{HJM11}.

The galaxy 4-velocity field is $u^\mu=dx^\mu/dt$. The past-pointing photon 4-momentum is $k^\mu=dx^\mu/dv$, where $v$ is the null affine parameter with $v=0$ at $O$. Then
 \be \label{zkdef}
1+z=u_\mu k^\mu, ~~ k^\mu=(1+z)(-u^\mu+n^\mu), ~~ u_\mu n^\mu=0,~ n_\mu n^\mu=1\,,
 \ee
where $n^\mu$ is a unit vector along the line of sight. For observers co-moving with the matter, an increment $dv$ in null affine parameter corresponds to a time increment $dt$, where
 \be \label{tvz}
dt=-u_\mu k^\mu dv=-(1+z)dv \,.
 \ee
We need to relate $v$ to $z$ by \eqref{zkdef}:
 \be \label{dzdvgen}
\frac{dz}{dv} = k^\nu \nabla_\nu (u_\mu k^\mu)= k^\mu k^\nu \nabla_\mu u_\nu \,,
 \ee
where the last equality follows since $k^\mu$ is a geodesic. The covariant derivative is split as
 \be\label{nabu}
\nabla_\mu u_\nu= \frac{1}{3}\Theta h_{\mu\nu}+ \sigma_{\mu\nu}+\omega_{\mu\nu}-u_\mu \dot{u}_\nu\,,
\qquad h_{\mu\nu}=g_{\mu\nu}+u_\mu u_\nu \,,
 \ee
where $h_{\mu\nu}$ projects into the galaxy instantaneous rest space, the dot indicates $u^{\mu} \nabla_{\mu}$ , $\Theta$ is the volume expansion rate ($\Theta=3H$ in a Friedmann model), $\sigma_{\mu\nu}$ is the shear, $\omega_{\mu\nu}$ is the vorticity and $\dot{u}_\mu$ is the acceleration. Now we will assume that the Universe is dominated by pressure-free matter, whereby $\dot{u}_\mu=0$. Putting everything together, we get
 \be \label{dzdv}
\frac{dz}{dv}=(1+z)^2 \left[ \frac{1}{3} \Theta
+ \sigma_{\mu\nu} n^\mu n^\nu \right].
 \ee
Now we integrate along the lightray from $O$ to $E$, using \eqref{tvz} and \eqref{dzdv}:
 \ba \label{lbt2}
t_O-t_E= \int_0^{z_E} \frac{dz}{(1+z) \Big[ \Theta(z)/3 + \sigma_{\mu\nu}(z) n^\mu n^\nu \Big]}\,.
 \ea
This will give us the look-back time -- {\em provided that we can uniquely relate the time intervals along galaxy worldlines that cross the lightray to a time interval along our worldline} ${\cal O}$. In order to do this, we need the existence of spatial 3-surfaces that are everywhere orthogonal to $u^\mu$; these will then be surfaces of constant proper time. The necessary and sufficient condition for these surfaces to exist is an irrotational flow:\footnote{This condition is only required on scales where the dust model holds: it is violated on nonlinear scales due to multi-streaming and baryonic effects.}
 \be
\omega_{\mu\nu}=0\,.
 \ee
Then we can uniquely identify the event $E'$ where the constant proper time surface $t=t_E$ through $E$ intersects ${\cal O}$. For rotating matter, it is not clear whether we can consistently define a look-back time.  From now on we assume that the general spacetime has irrotational cold matter and dark energy whose  perturbations are negligible, together with standard baryonic and radiation content.

A clear target of observational cosmology should thus be to {\em measure} $\Theta(z)$ and $\sigma_{\mu\nu}(z)$ in order to probe homogeneity. In order to identify the line of sight and transverse expansion rates in a general spacetime, we start from the matter expansion tensor 
\begin{equation}
\Theta_{\mu \nu} = \frac{1}{3} \Theta h_{\mu \nu} + \sigma_{\mu \nu}\,,\qquad {\Theta =  \Theta_{\mu \nu} h^{\mu \nu}}\,,
\end{equation}
where the average expansion rate is ${\frac{1}{3} \Theta}$.
The line of sight (radial) expansion rate is
\begin{equation}
{H_{\|}} := \Theta_{\mu \nu} n^{\mu} n^{\nu} = \frac{1}{3} \Theta + \sigma_{\mu \nu} n^{\mu} n^{\nu}\,, 
\end{equation}
so that the lookback time from \eqref{lbt2} is
 \ba \label{lbt}
t_O-t_E= \int_0^{z_E} \frac{dz}{(1+z) H_\|(z,n^\mu)}\,.
 \ea
The transverse expansion tensor is
\begin{equation}\label{thetat}
{\Theta_{\mu \nu}^{\perp}} = \Theta_{\alpha \beta} S^{\alpha}_{\mu} S^{\beta}_{\nu} = \frac{1}{3}{ \Theta_{\perp}} S_{\mu \nu} + \sigma^{\perp}_{\mu \nu}\,,
\qquad S_{\mu \nu} = h_{\mu \nu}-n_{\mu} n_{\nu}\,,
\end{equation}
where $S_{\mu \nu}$ is  the projector into the transverse space (``screen space"). 
Then the transverse expansion rate is
\begin{equation}
H_{\perp} =\frac{1}{2} \Theta_{\perp} = \frac{1}{2} \Theta^{\perp}_{\mu \nu} S^{\mu \nu} =  \frac{1}{2} \Theta_{\mu \nu} S^{\mu \nu}= \frac{1}{3} \Theta - \frac{1}{2} \sigma_{\mu \nu} n^{\mu} n^{\nu}\,. 
\end{equation}
Then it follows that, as required, the volume expansion rate is
\begin{equation}
\Theta(z)=H_{\|} (z, n^{\mu}) + 2 H_{\perp} (z, n^{\mu}) \,, 
\label{eq:exp}
\end{equation}
while the radial shear is
\ba \label{rads}
\sigma_{\mu\nu}(z) n^\mu n^\nu = \frac{2}{3} \Big[H_{\|} (z, n^{\mu}) - H_{\perp} (z, n^{\mu})\Big]  \,.
\ea
The shear can be split into transverse, radial and mixed parts:
\ba\label{sig1}
\sigma_{\mu\nu}= \sigma^\perp_{\mu\nu} + A n_\mu n_\nu + 2 B_{(\mu}n_{\nu)}\,,~~~ B_\mu n^\mu=0\,, 
\ea
where $\sigma^\perp_{\mu\nu}$ is defined by \eqref{thetat},
and  $A,B_\mu$ are found by suitable contractions of \eqref{sig1}. This leads to
\ba \label{sig2}
\sigma_{\mu\nu}= \sigma^\perp_{\mu\nu}+ \frac{2}{3} \big(H_{\|}-H_\perp  \big) n_\mu n_\nu+ 2 \sigma_{\alpha\beta}n^\alpha S^\beta_{(\mu}n_{\nu)}\,.
\ea
In principle, $H_{\|}$ is determined by baryon acoustic oscillation (BAO) measurements of a physical, radial length -- a standard ``ruler'' -- in galaxy clustering \cite{BAO}: 
\ba
{{H_\|}=\frac{c}{(1+z) \Delta r_\|} \, \Delta z}\,,
\ea
or by cosmic chronometers using a standard ``clock'' in the form of differential ages of ancient, elliptical galaxies \cite{CC,simon}: 
\begin{equation}
{ H_{\|} = -\frac{ \Delta z}{(1+z) \Delta t}} \,,
\end{equation}
which follows from \eqref{lbt}. However, while the cosmic chronometer method is fully independent of the cosmological model, the radial length BAO needs to assume a value for $\Delta r_\|$ or obtain it through consistency with other measurements. These are the only\footnote{Observations of supernova as standarizable candles also give $H_{\|}$ but this depends on assuming the metric.} two routes to obtain $H_{\|}$.

Once $H_{\|}$ is determined, we would be able to find $H_\perp$ if we could probe  the remote volume expansion $\Theta$, using \eqref{eq:exp}. By \eqref{rads} or \eqref{sig2}, an alternative would be available if we could probe the remote shear $\sigma_{\mu\nu}$. Then we would be able to  test homogeneity by testing isotropy of the expansion rate at remote locations. The problem is to find a direct observational way to determine $H_\perp$ or $\Theta$ or $\sigma_{\mu\nu}$. In the absence of a direct solution, we turn to investigate the information contained in the evolution of polarization.

\section{Polarization  in a general cosmological spacetime}\label{gmet}

Polarization in a perturbed Friedmann model is well understood (see e.g.~\cite{Arthur,Zaldarriaga:1996xe,Kamionkowski:1996ks}). Linear polarization is described by the Stokes parameters $Q, U$. Note that these parameters have units of intensity per unit frequency, their measurement inevitably involve a quantity that is an integration of these parameters over a frequency range. In this sense the  $Q, U$ parameters should be  seen as ``differential" quantities.
Under rotations through $\phi$ in the screen space, these parameters $Q',U'$ are rotated by $2\phi$ in parameter space, showing that linear polarization is described invariantly by a spin-2 object in the screen space. Thus $Q,U$ are not physical invariants but depend on coordinates in the screen space. The invariants under rotation are
\ba \label{prot}
Q^{\prime 2}+U^{\prime 2} =Q^2+U^2  \,,
\ea
whereas the direction defined by
the polarization angle,
\ba
\alpha \equiv \frac{1}{2} \tan^{-1}\frac{U}{Q}~\Rightarrow~ \alpha' =\alpha-\phi\,
\ea
is not invariant.

In a general cosmological spacetime (i.e. without assuming a background or any large-scale symmetries), we need to deal with the invariant objects. A general analysis was developed in a pioneering paper by Challinor~\cite{Challinor:2000as} (see also~\cite{Anile,Tsagas:2007yx}): linear polarization is described by a symmetric  trace-free tensor ${\cal P}^{\mu\nu}$ in the screen space, i.e. a spin-2 object in the screen space, which satisfies 
\ba
{\cal P}^{\mu\nu}S_{\mu\nu} =0={\cal P}^{\nu\mu}-{\cal P}^{\mu\nu}\quad \mbox{and}~~
{\cal P}^{\mu\nu}n_\nu =0={\cal P}^{\mu\nu}u_\nu~\big(\mbox{or}~~ {\cal P}^{\mu\nu} ={\cal P}^{\mu\nu}_\perp := S^\mu_\beta S^\nu_\gamma\, {\cal P}^{\beta\gamma}\big).
\ea
The magnitude of the polarization tensor is independent of coordinate choice and is given by the rotational invariant \eqref{prot}~\cite{Challinor:2000as}:
\ba \label{pmag}
2\,{\cal P}_{\mu\nu}\,{\cal P}^{\mu\nu} = Q^2 + U^2\,.
\ea
After scattering by free electrons in a scatterer located at a given redshift $z$ which is composed of a collapsed dark matter halo above a mass large enough to host high-energy free electrons that cause inverse Compton scattering on lower energy CMB photons, the linear polarization tensor in the screen space propagates along lightrays towards the observer according to conservation of $\nu^{-3}\, {\cal P}^{\mu\nu}$, where $\nu$ is the photon frequency~\cite{Challinor:2000as}:
\ba \label{fpal}
\big[k^\alpha\nabla_\alpha\big(\nu^{-3} \,{\cal P}^{\mu\nu}\big)\big]_\perp = S^\mu_\beta S^\nu_\gamma \, k^\alpha\nabla_\alpha\big(\nu^{-3}\, {\cal P}^{\beta \gamma}\big)= 0\,.
\ea
Note that we do not impose the stronger condition $k^\alpha\nabla_\alpha(\nu^{-3} \,{\cal P}^{\mu\nu}) = 0$, since in general lightray derivatives of screen-space quantities do not lie purely in the screen space. Polarization measurements implicitly involve a projection into the screen space, so that any components not in the screen space do not affect the measurement.

If we project \eqref{fpal} with $\nu^{-3} \,{\cal P}_{\mu\nu}$, we have
\ba
0=k^\alpha\nabla_\alpha\big[\nu^{-3} \,{\cal P}_{\mu\nu}\big(\nu^{-3} \,{\cal P}^{\mu\nu}\big)\big]
=\frac{1}{2}k^\alpha\nabla_\alpha\big[\nu^{-6}\big(Q^2+U^2\big)\big]\,.
\ea
It follows that for a source $E$  observed by $O$ at redshift $z=\nu_E/\nu_O-1$, we have
\ba \label{quc}
Q_E^2+U_E^2= (1+z)^6 (Q_O^2+U_O^2)\,.
\ea
This is the expected scaling with redshift for the differential Stokes parameters.

\subsection{Local coordinates for polarization}
 
 For matter that is irrotational and pressure-free on large scales, we have
 \be
 \omega_{\mu\nu}=0=\dot{u}_\mu ~~\Leftrightarrow ~~ u_{[\mu,\nu]}=0 ~~\Leftrightarrow ~~  u_{\mu}=-t_{,\mu}\,, \ee
for some scalar $t$ -- which is then necessarily the proper time along matter worldlines. Therefore we can choose comoving coordinates $(t,x^i)$ such that
\ba
ds^2 &=& g_{\mu\nu}dx^\mu dx^\nu=-(u_\mu dx^\mu)^2+h_{\mu\nu}dx^\mu dx^\nu \nonumber \\
&=& -dt^2+(n_idx^i)^2+S_{ij}dx^i dx^j\,.
\ea
Locally, i.e., in a neighborhood of any point, we can choose $x^1=x$ along $n^i$ and then 
\be
ds^2\big|_{\rm loc}=-dt^2+ A_\|^2 d x^2+S_{IJ}\,d x^I d x^J\,,
\ee 
where $ x^I=( y, z)$ and $ n_i=A_\|\delta_i^1$. 
The area element in the screen space is  $dV_\perp= \sqrt{\det S_{IJ}}\,d^2x$. Transverse areas expand as
$A_\perp^2$, where $A_\perp$ is the transverse scale factor; since $x^I$ are comoving (constant along the matter world-lines) this means that $\sqrt{\det S_{IJ}}\propto  A_\perp^2$. We can normalize $A_\perp$ at some time $t=t_0$ so that
$\sqrt{\det S_{IJ}}=A_\perp^2$, and then $S_{IJ}=A_\perp^2 s_{IJ}$, where $\det s_{IJ}=1$. Thus
\be  \label{polco}
ds^2\big|_{\rm loc}=-dt^2+ A_\|^2 d x^2+A_\perp^2\,{s_{IJ}}\,d x^I d x^J \quad\mbox{where}\quad \det s_{IJ}=1\,.
 \ee
The expansion rates are  
\be\label{ExpansionRates}
H_\|={\dot{A}_\| \over A_\|}\,, \quad H_\perp={\dot{A}_\perp \over A_\perp}\,.
\ee
{Note that $H_\perp$ is the geometric mean of the expansion rates in the local principal axis system of $S_{IJ}$}.
 
In these coordinates, the polarization tensor has only screen-space components, and these components are the Stokes linear polarization parameters $Q,U$ measured by the observer using the local coordinates in the screen space:
\be\label{pij}
{\cal P}_{\mu\nu}= {\cal P}_{IJ}\,\delta^I_\mu\, \delta^J_\nu\,,\quad {\cal P}_{IJ}= {1\over 2} \left( \begin{array}{cc} Q & ~~U \\ U & -Q \end{array} \right).
\ee
We used  $S^I_\mu=S^I_J \delta^J_\mu$ and $S^I_J=\delta^I_J$, which hold in the polarization coordinates of \eqref{polco}.

An alternative to local coordinates is an orthonormal tetrad. A polarization tetrad is briefly described in Appendix A.

\subsection{Drift of polarization}

The time evolution of polarization at a scatterer is given in a general spacetime by the covariant derivative  of the polarization tensor along the four-velocity of the scatterer, projected into the screen space, i.e. by $\big(\dot {\cal P}_{IJ}\big)_\perp$ at $E$. 
In the local coordinates of \eqref{polco}, both $S_{\mu\nu}$ and ${\cal P}_{\mu\nu}$ are zero if $\mu$ or $\nu$ is 0 or 1, and we find that
\ba\label{pdot}
\big(\dot {\cal P}_{IJ}\big)_\perp :=S^\mu_I S^\nu_J\, \big(u^\alpha\nabla_\alpha\, {\cal P}_{\mu\nu}\big) = {\cal P}_{IJ,0} - 
\Gamma^K_{I0}\,{\cal P}_{KJ} - \Gamma^K_{J0}\,{\cal P}_{IK}\,.
\ea
The Christoffel symbols in \eqref{pdot} encode the screen-space shear and the volume expansion rate:
\ba \label{gam}
\Gamma^K_{I0} = \sigma_{\perp\,I}^{\,K}+{1\over3}\big(H_\|+2H_\perp \big)\delta^K_I\,.
\ea
This can be seen as follows. By  \eqref{nabu}, with $\omega_{\mu\nu}=0$ and $u^\mu=\delta^\mu_0$, we have 
\ba \label{sigma}
\sigma^\mu_\nu= \nabla_\nu u^\mu-{1\over3}\Theta\, \delta^\mu_\nu= \Gamma^\mu_{\nu 0}-{1\over3}\Theta \,\delta^\mu_\nu \,.
\ea
Then we use \eqref{eq:exp} for $\Theta$ and project into the screen space to obtain \eqref{gam}. 
We can rewrite \eqref{pdot} as
\ba\label{pdot2}
\big(\dot {\cal P}_{IJ}\big)_\perp = {d\over dt} {\cal P}_{IJ} - {2\over3}\big(H_\|+2H_\perp \big){\cal P}_{IJ}-2
\sigma_{K(I}^\perp\,{\cal P}_{J)}^K\,.
\ea
This equation can be derived also using the tetrad in the Appendix without any need to use local coordinates. By the Equivalence Principle, $d {\cal P}_{IJ} /dt$ is given by the special relativistic scattering formula, which depends on the properties of the free electron distribution in the scatterer and of the CMB photons, both of which can be estimated from observations. The observable $(\dot {\cal P}_{IJ})_\perp$ is therefore determined by the local scattering physics (via $d {\cal P}_{IJ} /dt$) and by gravitational effects, which produce the expansion rate $(H_\|+2H_\perp)/3$ and screen-space shear $\sigma^\perp_{IJ}$, of the matter field.

If we observe a scatterer over a proper time interval $\delta t_O$ at the observer, where
\ba\label{td}
\delta t_O=(1+z)\delta t_E\,,
\ea 
then it follows from \eqref{quc}
that the change in polarization magnitude at the scatterer is related to the observed change in polarization magnitude by
\be \label{pold1}
\delta \big(Q^2+U^2\big)_E = (1+z)^{6}\,\delta \big(Q^2+U^2\big)_O+ 6(1+z)^{5} \big(Q^2+U^2\big)_O\,\delta z\,,
\ee 
where the redshift measured at the observer is $z+\delta z$.

Equation \eqref{pold1} {\em predicts the polarization drift at the scatterer in terms of the measured polarization drift and redshift drift at the observer.} The polarization drift at the scatterer is also determined by \eqref{pdot2}:
\ba\label{dpeqn}
\delta {\mathcal{P}_{IJ}}\big|_E = \big(\dot {\cal P}_{IJ}\big)_{\perp\,E} \delta t_E \,,
\ea
where $\delta t_E$ is the proper time interval at the scatterer and $(\dot {\cal P}_{IJ})_{\perp\,E}$ is given by \eqref{pdot2}.
By comparing the theoretical prediction for the polarization drift with the measurement \eqref{pold1}, we can in principle deduce the local volume expansion rate and the screen-space shear at the scatterer. If we  also find the radial expansion rate via the BAO, then we can deduce the transverse expansion rate at the scatterer. To be more specific, from local measurements of $z$ and $\delta t_O$ we can obtain $\delta t_E$ as in \eqref{td}. From measurements of the redshift drift $\delta z$ (which can be measured directly from estimates of $H_0$ using the local distance ladder), ${\cal P}_O$ and $\delta{\cal P}_O$ we can use \eqref{pold1} and \eqref{dpeqn} to determine $\big(\dot {\cal P}_{IJ}\big)_{\perp\,E}$. Then we use the two equations in \eqref{pdot2} and supply a theoretical prediction for ${d} {\cal P}_{IJ}/dt$ to get $H_\perp$ and $\sigma^\perp_{IJ}$. This is our procedure to measure homogeneity.

\section{Observational Strategy}

We can provide an estimate of the observations needed to constrain homogeneity with the method developed above. It is beyond the scope of this paper to provide a detailed study of the experimental setup needed: this will be presented elsewhere. 

Our  proposed method relies on the difficult task of measuring the polarization drift, i.e.,  the time variation of the polarization tensor, at each scatterer position. The redshift drift (see Appendix B) needs knowledge of $H_0$ which has already been obtained at the \% level with the local distance ladder and the other relevant  quantities are much  easier to measure and have been discussed extensively in the literature.  Effectively, one needs to ``film''  polarization (for a closely related idea see also~\cite{Lyman}; also see \cite{Zibin:2007mu,Moss:2007bu}). 

While the polarized cosmological signal can be found in several observables, we seek a combination of detection method, observable and its scatterer that achieves the following:
\begin{enumerate}
\item It is stable enough to be observed for a long time and thus to detect small drifts. 
\item The polarization  signal can be measured with exquisite signal to noise.
\item  The scatterer is at cosmological distances and  its  redshift   can be reliably measured (this does not need to be spectroscopic but can be photometric, which already exist).
\item It is abundant. 
\item The signal  can be easily accessible with current technology (but not necessarily with current experiments). 
\end{enumerate} 

For this reason we focus on the polarized signal of  CMB photons that have been inverse-Compton scattered by the hot intra-cluster gas of massive galaxy clusters. Consider a radio telescope with spatial resolution at the $\sim$ arcmin level. This is achievable  as   it is not too dissimilar to that of  the  Planck space mission.  Consider also that measurements can be obtained over the time frame of ${\cal O}(10)$ years and that future CMB polarization experiments will be basically photon-noise limited because of the large number of detectors on the focal plane. 

Halos of  dark matter mass  above  $10^{13}$ M$_{\odot}$ are optimal scatterers, leaving their easily  identifiable (Sunyaev-Zel'dovich~\cite{SZ}) signature  on  CMB high-resolution maps. An experiment to detect this signal is something like the more updated versions of CMB-S4 \cite{Abitbol:2017nao}  considered by Ref.~\cite{wandelt}, ($N_{\rm det} = 10^7$ detectors, $D=12m$ mirror).  Since the drift is linear in time, there is a considerable gain through having a longer experiment, with the error on the rate decreasing as $t_{\rm exp}^{-3/2}$. For a mission with improved detector sensitivitiy $s_{\rm det}$, from the CoRE proposal
\footnote{\url{http://www.core-mission.org/documents/CoreProposal_Final.pdf}}
, with a baseline $1.2m$ mirror, and mission length of $\delta t= 4 yr$, the noise level is 
\begin{equation}
c_{\rm noise} = 4.7 \mu K {\rm arcmin} \,\left(\frac{4 yr}{\delta t}\right)^{1/2}\,
\left(\frac{400}{N_{\rm det}}\right)^{1/2}\,\left(\frac{s_{\rm det}}{50\mu K s^{1/2}}\right)
\end{equation}
The S/N on the normalised drift rate $a$, defined such that the polarisation signals evolve from the initial observation $P_0$ at $t=0$
\begin{equation}
P(t) = P_0\left (1+ a \frac{t}{t_*}\right)
\end{equation}
(where $t_*$ is the expansion timescale) is obtained through a Fisher analysis of the error on $a$, which yields an error
\begin{equation}
\sigma_a = \frac{\sqrt{6}}{N_{\rm pix}^{1/2}S}\left(\frac{\delta t \,t_*^2}{t_{\rm exp}^3}\right)^{1/2}
\end{equation}
where $N_{\rm pix}$ is the number of pixels in the polarisation map, which we assume is repeatedly measured once every $\delta t$.  Putting these together, assuming all-sky coverage, the signal-to-noise for the polarisation drift would make a detection challenging with the following $S/N$:
\begin{equation}
\frac{S}{N} = 66 \left(\frac{N_{\rm det}}{10^7}\right)^{1/2} \,\left(\frac{D}{12m}\right) \, \left(\frac{s_{\rm det}}{0.1\,\mu K s^{1/2}}\right)^{-1}
\, \left(\frac{t_{\rm exp}}{{10\ yr}}\right)^{3/2} \, \left(\frac{t_*}{{\ Gyr}}\right)^{-1}  
\end{equation}
Foreground variations are likely to be uncorrelated with the drift, but would constitute an additional source of noise. As pointed out in~\cite{wandelt}, the main contaminant is the $E$ primordial mode. Our task is,  on the other hand, easier as  we only need to measure differential  variations, which minimizes greatly  many systematic effects\footnote{Stacking galaxy clusters in the same redshift slice will eliminate any intrinsic variations in the cluster evolution.}.  Thus it is not unreasonable to assume that our differential measurement could have a S/N of ${\cal O} (100)$  in the integrated full sky. Assuming scatterers can all be identified in CMB maps and assuming the Stokes parameters can be reliably measured for all of them, we could limit variations of  $H_{||} + 2H_{\perp}$ via \eqref{pold1} and \eqref{pdot2}.
 
Recall that we need to measure $H_{||}$ independently of the metric to determine $H_{\perp}$. The BAO technique does require a value of the sound horizon that is usually assumed to be the one given by the CMB, which assumes homogeneity even when using only local measurements to obtain the ruler's length \cite{ruler}. On the other, hand none of these assumptions are needed for the cosmic chronometer method, that is fully independent of the metric of space-time or the cosmological model. For the sake of the argument here we can assume that in future measurements $H_{||}$ can be measured at the percent level. This will be the degree that we can constrain homogeneity with future surveys. However, it is worth recalling that the Planck space mission already has observed  $10^3$ Sunyaev-Zel'dovich clusters for which the polarization drift could, in principle, be measured. This could give an interesting constraint on the degree of homogeneity; we will explore this  elsewhere. We  are fully aware that we have ignored many real-world effects, like foregrounds and other intrinsic time variable effects on $Q$ and $U$, but we have shown that the method to measure homogeneity  presented  above  is,  in principle,  feasible.

\section{Conclusions} 

Measuring the degree of homogeneity of the space-time metric of the Universe remains an open question in cosmology. We have presented a method to measure homogeneity in general space-time metrics by ``filming" the polarization signal  of  CMB photons inverse Compton scattered by the hot intra-cluster gas in galaxy clusters. In particular, the change in time of the Stokes parameters provides a measurement of the transverse expansion rate. The radial expansion rate is instead measured by more conventional probes like radial BAO or cosmic chronometers. We have estimated  that a measurement of homogeneity at the $\sim$ percent level can be obtained with high resolution full sky CMB polarization maps in a period of years. Percent-level constraints on the degree of homogeneity may be achievable with the  expected sensitivity of the proposed Simons Observatory \cite{Ade:2018sbj} and CMB-S4 experiment \cite{Abazajian:2016yjj}.

\[ \]{\bf Acknowledgments:} We thank Anthony Challinor, Chris Clarkson and Julien Larena for helpful discussions. RJ and LV thank the Center Emile Borel for hospitality during the latest stages of this work. Funding for this work was partially provided by the Spanish MINECO under projects AYA2014-58747-P AEI/FEDER, UE, and MDM-2014-0369 of ICCUB (Unidad de Excelencia Mar\'ia de
Maeztu).  LV acknowledges support from the  European Union Horizon 2020 research and innovation program
ERC (BePreSySe, grant agreement 725327). RM acknowledges support from the South African SKA Project and the National Research Foundation of South Africa (Grant No. 75415). RM was also supported by the UK Science \& Technology Facilities Council (Grant No. ST/N000668/1). The work of RC is supported in part by the US Department of Energy grant DE-SC0010386.

\appendix

\section{Polarization tetrad}

An orthonormal tetrad $\bm{e}_a=(\bm{u},\bm{n},\bm{e}_A)$, where $\bm{e}_A$ are orthogonal unit vectors spanning the screen space, is adapted to describe polarization, which is measured in the screen space by an observer $\bm{u}$. The tetrad components ${\cal P}_{ab}= {\cal P}_{\mu\nu}\,e_a^\mu\, e_b^\nu$ are then physical quantities. In this tetrad, the polarization tensor has nonzero components only in the screen space, and these components define the linear polarization Stokes quantities $\hat{Q},\hat{U}$ that are measured by the observer:
\be\label{pol0}
{\cal P}_{AB}\equiv {\cal P}_{\mu\nu}\,e_A^\mu\, e_B^\nu = {1\over 2} \left( \begin{array}{cc} \hat{Q} & ~~\hat{U} \\ \hat{U} & -\hat{Q} \end{array} \right).
\ee
We use hats to distinguish the Stokes parameters in the polarization tetrad from those in the polarization coordinates of \eqref{pij}.

The orthonormal tetrad $\bm{e}_a$ has rotational freedom in the the screen-space basis $\bm{e}_A$. By \eqref{fpal}, a natural choice  is  that $\bm{e}_A$   propagates along the lightrays  according to
\be \label{fert}
\big(k^\alpha\nabla_\alpha e_A^{\mu}\big)_\perp =0\,.
\ee
With this choice of the screen-space basis -- which we can call the polarization basis -- it follows from \eqref{fpal} that the tetrad components ${\cal P}_{AB}$
propagate according to
\be\label{pol1}
{d\over dv}\big(\nu^{-3}\, {\cal P}_{AB}\big) =0\quad \mbox{equivalently}\quad  {d\over dv}\big(\nu^{-3}\,\hat{Q}\big)=0=
{d\over dv}\big(\nu^{-3}\,\hat{U}\big)\,.
\ee

A consequence of \eqref{pol1} is that the polarization at the scatterer is given in terms of the polarization measured at the observer by
\be\label{pol2}
\big(\hat{Q}_E\,,\, \hat{U}_E\big) = (1+z)^{3}\, \big(\hat{Q}_O\,,\, \hat{U}_O\big)\,,
\ee
where $z$ is the observed redshift of the scatterer.
In particular, this means that the polarization angle $\alpha$ is constant along each lightray:
\be\label{pol3}
\tan 2\hat{\alpha} \equiv {\hat{U}\over \hat{Q}} ~~\Rightarrow~~ {d\hat{\alpha}\over dv }=0~~\Rightarrow~~ \hat{\alpha}_E=\hat{\alpha}_O\,
\ee
Note that \eqref{pol1}--\eqref{pol3} hold only in the polarization tetrad defined by \eqref{fert}.

\section{Redshift drift in a general cosmological spacetime}

It follows from (2.2)  and (2.5) that
\ba
1+z=\exp \int_{t_E}^{t_O} dt\,H_\|(t,n^\mu)
\ea
Consider the small change $\delta z$ in $z$ over a proper time interval $\delta t_O$ at the observer. The corresponding time interval along the $u^\mu$ world-line at the source is $\delta t_E$, and
\ba
(1+z+\delta z)-(1+z)= \exp \int_{t_E+\delta t_E}^{t_O+\delta t_O}dt\,H_\|-\exp\int_{t_E} ^{t_O} dt\,H_\|
\ea
We break up the total time interval $t_E\to  t_O+\delta t_O$ into 4 segments, 
\ba
\delta z
&=&  \exp\Big( \int_{t_E+\delta t_E}^{t_O}dt\,H_\| + \int_{t_O}^{t_O+\delta t_O}dt\,H_\|\Big)
-\exp\Big(  \int_{t_E}^{t_E+\delta t_E}dt\,H_\|+\int_{t_E+\delta t_E} ^{t_O}dt\,H_\|\Big)  \nonumber\\
&=&  \Big( \exp \int_{t_E+\delta t_E}^{t_O} dt\,H_\| \Big) \times \Big[\exp \int_{t_O}^{t_O+\delta t_O}dt\,H_\|
-\exp\int_{t_E}^{t_E+\delta t_E}dt\,H_\|\Big]
\ea
Now $\delta t_E=(1+z)^{-1}\delta t_O$, and  working to lowest order in $\delta t_O$:
\ba
\delta z &\approx& 
 \Big( \exp \int_{t_E}^{t_O} dt\,H_\| \Big) \times \Big[\exp \int_{t_O}^{t_O+\delta t_O}dt\,H_\|
-\exp\int_{t_E}^{t_E+\delta t_E}dt\,H_\|\Big]  \nonumber\\
&\approx& 
 (1+z) \Big\{\exp\big[\delta{t_O} H_\|(t_O,n_O^\mu)\big]
-\exp\big[\delta{t_E} H_\|(t_E,n_E^\mu)\big] \Big\}  \nonumber\\
&\approx& 
 (1+z) \Big[1+\delta{t_O} H_\|(t_O,n_O^\mu)
-1-(1+z)^{-1}\delta{t_O} H_\|(t_E,n_E^\mu) \Big]  
\ea
Finally
\ba
{\delta z\over \delta t_O} = (1+z)H_\|(t_O,n_O^\mu)-H_\|(t_E,n_E^\mu) +O(\delta t_O^2)
\ea

\end{document}